\documentclass[twocolumn,10point,amsmath,amssymb]{revtex4}

\usepackage{graphicx} 
\usepackage{bm} 

\begin{document}
\newcommand{\ethaffil}{Laboratory of Physical Chemistry and optETH, ETH Zurich, CH-8093 Zurich, Switzerland}
\newcommand{\real}{\operatorname{Re}}

\title{Extinction imaging of a single quantum emitter in its bright and dark states}

\author{P. Kukura}
\author{M. Celebrano$^{\dag }$}
\author{A. Renn}
\author{V. Sandoghdar}\email{vahid.sandoghdar@ethz.ch}

\affiliation{Laboratory of Physical Chemistry and optETH, ETH
Zurich, CH-8093 Zurich, Switzerland\\
$^{\dag }$Permanent address: Department of Physics, Politecnico di
Milano, 20133 Milan, Italy}

\maketitle

\textbf{Room temperature detection of single quantum emitters has
had a broad impact in fields ranging from
biophysics~\cite{Betzig:06,Michalet:05} to material
science~\cite{Kirstein:07}, photophysics~\cite{Dickson:97}, or even
quantum optics~\cite{Michler:00}. These experiments have exclusively
relied on the efficient detection of fluorescence. An attractive
alternative would be to employ direct absorption, or more correctly
expressed ``extinction''~\cite{Bohren-83book} measurements. Indeed,
small nanoparticles have been successfully detected using this
scheme in reflection~\cite{Lindfors:04,Ignatovich:06,Ewers:07} and
transmission~\cite{Mikhailovsky:03,Arbouet:04}. Coherent extinction
detection of single emitters has also been reported at cryogenic
temperatures~\cite{Plakhotnik:01,Karrai:03,Gerhardt:07a,Wrigge:08},
but their room temperature implementation has remained a great
laboratory challenge owing to the expected weak signal-to-noise
ratio~\cite{Hwang:06}. Here we report the first extinction study of
a single quantum emitter at ambient condition. We obtain a direct
measure for the extinction cross section of a single semiconductor
nanocrystal both during and in the absence of fluorescence, for
example in the photobleached state or during blinking
off-times~\cite{nirmal:96}. Our measurements pave the way for the
detection and absorption spectroscopy of single molecules or
clusters of atoms even in the quenched state.}

Ensembles of emitters, such as atoms, molecules, ions, or quantum
dots, are routinely studied via absorption or fluorescence
spectroscopy alike. However, conventional absorption spectroscopy is
difficult to perform on very dilute samples because one has to
detect very small changes on a large signal. For a dye molecule
placed in a diffraction-limited laser spot, a simple estimate
predicts an extinction dip of about $2 \times 10^{-6}$ if we
consider a typical cross section of $\sigma \sim 1 \times
10^{-15}~\rm cm^{2}$ at room temperature. In this letter we
demonstrate that a direct extinction measurement can be nevertheless
successful in detecting a single quantum emitter without using any
noise suppression methods such as lock-in detection. We performed
our experiments on semiconductor nanocrystals, which have been shown
to behave like artificial atoms with well-defined quantized energy
levels and nonclassical emission photon
statistics~\cite{Michler:00}. We chose such nanocrystal quantum dots
(NQD) because they have slightly larger absorption cross sections
than dye molecules and last much longer than these before
photobleaching~\cite{Peng:97}.

\begin{figure}[b!]
\centering
\includegraphics[width=8 cm]{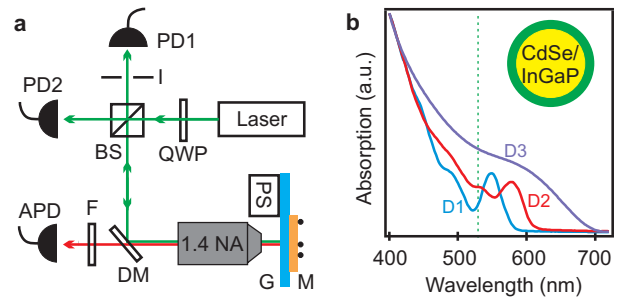}
\caption{\textbf{Setup}. (a) Schematics of the optical system and
sample arrangement. PD1: photodiode for detection of the extinction
signal; PD2: photodiode for normalization of the laser intensity; I:
Iris; BS: 50:50 beam splitter; QWP: quarter wave plate; DM: dichroic
mirror; F: long pass filter; APD: avalanche photodiode; G:
microscope cover glass slip; M: mica sheet; PS: piezo scanner. (b)
Ensemble extinction spectra of the nanocrystal quantum dots used in
this work. Details are provided in the Methods section. The curves
are scaled to match at the short wavelength end.}\label{setup}
\end{figure}

The schematics of our experimental setup is depicted in
Fig.~\ref{setup}a and described in the Methods section. Here it
suffices to state that we illuminate the sample with a focused laser
beam at a wavelength of 532~nm and detect the fluorescence and the
reflected excitation light on separate detectors. Images are
produced by raster scanning the sample. We examined various
commercially available NQDs with ensemble extinction spectra shown
in Fig.~\ref{setup}b. The NQDs were spin cast on freshly cleaved
thin mica sheets that are known to yield locally atomically flat
surfaces~\cite{Hegner:93}. This was important for minimizing the
fluctuating background that one encounters when employing the scheme
of Fig.~\ref{setup}a on a standard microscope cover
glass~\cite{Jacobsen:06}.

\begin{figure}[b!]
\centering
\includegraphics[width=8.5 cm]{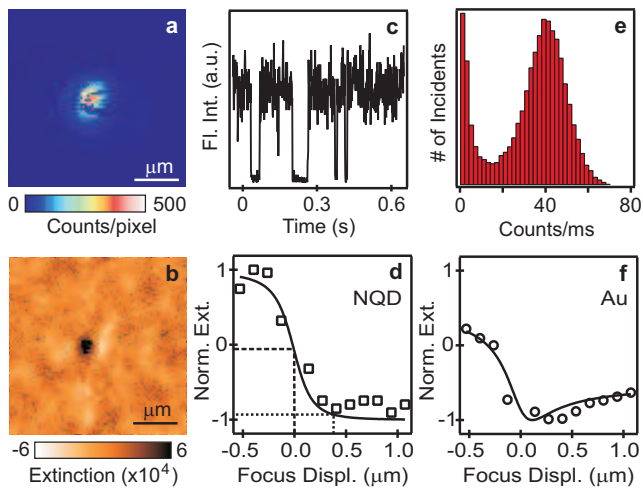}
\caption{\textbf{Extinction detection of a single quantum dot} (a) A
fluorescence image of an individual D2 quantum dot (excitation
intensity of 2~kW/$\rm cm^2$). (b) Corresponding extinction image (6
averages). (c) Partial fluorescence trajectory of the dot in (a)
(time bin: 1~ms, excitation intensity 200~W/$\rm cm^2$). (d) The
normalized variation of the extinction signal of a D3 type NQD as a
function of its displacement from the focus along the optical axis
(symbols: experimental; line: theoretical). Here we chose D3 dots
because their large $\sigma$ provided reliable signals even away
from the maximum. (e) Histogram of fluorescence counts collected
from the dot in (a) over 50~s. (f) Same as in (d) but for a single
gold nanoparticle with a diameter of 10~nm. }\label{detection}
\end{figure}

Figures~\ref{detection}a and b show examples of simultaneously
recorded fluorescence and reflection images of a sample prepared
using D2 NQDs. The observed intermittent fluorescence in (a) proves
that the signal stems from a single NQD~\cite{Dickson:97}.
Fig.~\ref{detection}c displays this more clearly by a time trace of
the fluorescence signal recorded from the NQD under constant
illumination. The on-off blinking behavior of fluorescence is
illustrated further in Fig.~\ref{detection}e by a histogram of the
fluorescence counts, yielding two well defined peaks. Thus,
Fig.~\ref{detection}b presents an image of a single quantum emitter
recorded in reflection with a signal contrast of about $6\times
10^{-4}$. This is substantially larger than the estimate we stated
earlier. We now explain the origin of this enhanced signal and
relate it to the extinction cross section of a NQD.

In a transmission experiment, the signal $I_{\rm det}$ on the
detector can be obtained by writing~\cite{Bohren-83book}
\begin{equation}
I_{\rm det}=\left\vert \rm E_{exc}+\rm E_{sca}\right\vert
^{2}=\left\vert \rm E_{exc}\right\vert ^{2}+\left\vert \rm E_{sca}
\right\vert ^{2}+2\rm Re[E^*_{exc}E_{sca}] . \label{signal}
\end{equation}
where $\mathbf{E}_{\rm exc}$ denotes the complex electric field of
the excitation beam. The field $\mathbf{E}_{\rm sca}$ scattered by
the sample is given by $\mathbf{E}_{\rm sca} \propto |\alpha|
e^{i\phi_{\rm sca}} \mathbf{E}_{\rm exc}$ where $\alpha$ is the
complex polarizability of the particle, and $\phi_{\rm sca}$ is the
scattering phase shift determined by the real and imaginary parts of
$\alpha$. The last term in Eq. (1) is known as the ``extinction
signal" and signifies the interference between $\mathbf{E}_{\rm
exc}$ and $\mathbf{E}_{\rm sca}$. For small particles it is much
larger than the second term and can be expressed as the sum of the
absorption and scattering cross sections~\cite{Bohren-83book}. In
reflection, $\mathbf{E}_{\rm exc}$ is replaced by $r\mathbf{E}_{\rm
exc}$ where $r$ is the sample reflectivity. Hence, the signal
contrast is increased by $r^{-1}\sim5$ for the mica/air
interface~\cite{Lindfors:04}. Measurements in reflection or
transmission are otherwise equivalent because in each case the
scattered light interferes with the excitation beam. Therefore, we
refer to our signal obtained in reflection also as the extinction
signal.

A second mechanism for enhancing the extinction signal in our
experiment stems from the optimization of the phase difference
between $\mathbf{E}_{\rm exc}$ and $\mathbf{ E}_{\rm sca}$. The
symbols in Fig.~\ref{detection}d show the measured variation in the
extinction signal of a NQD as a function of its displacement from
the focus. This strong position dependence is caused by the change
in the phase $\phi_{\rm exc}-\phi_{\rm sca}$ where $\phi_{\rm exc}$
denotes the phase of the focused laser beam dictated by the Gouy
phase~\cite{Hwang:07}. The solid line plots a theoretical fit.
Details are provided in the Methods section. This data reveal that
the extinction signal is enhanced by roughly ten times if one moves
away from the focus by +400~nm. Indeed, in each run we optimized the
signal by displacing the sample.

Obtaining a value for the extinction cross section based on the
measured contrast is not an easy matter because we do not have a
quantitative knowledge of the mode overlap between $\mathbf{E}_{\rm
exc}$ and $\mathbf{E}_{\rm sca}$~\cite{Gerhardt:07a}. To remedy this
difficulty, we chose to compare the signal contrast obtained from a
NQD with that of a gold nanoparticle (GNP) of diameter 10~nm, which
has a well-known extinction cross section~\cite{Bohren-83book}. We
prepared a sample containing both GNPs and NQDs and imaged them
under identical conditions, yielding an extinction contrast of
$3.5\times 10^{-3}$ for a single GNP. As shown by the symbols in
Fig.~\ref{detection}f, the extinction signal of a GNP does not vary
substantially within a displacement of +400~nm from the focus.

By comparing the in-focus extinction contrasts of $3.5\times
10^{-3}$ and $6\times 10^{-5}$ for GNPs and NQDs respectively, and
taking the literature value of $\sigma_{\rm Au} =2.7\times
10^{-13}~\rm cm^{2}$~\cite{Arbouet:04}, we arrive at $\sigma_{\rm
NQD} =4.5\times 10^{-15}~\rm cm^{2}$. This value is in good
agreement with $2\times 10^{-15}~\rm cm^{2}$ obtained from ensemble
extinction measurements~\cite{Leatherdale:02} and $3\times
10^{-15}~\rm cm^{2}$ extracted from fluorescence studies of similar
sized single NQDs~\cite{Kuno:01}. This level of agreement between
our measured values of extinction contrast and the expected
extinction cross sections holds for all three quantum dots in
Fig.~\ref{setup}b. However, we emphasize that we observe a
substantial spread of up to 3 times in the extinction signals
recorded from different individual NQDs of nominally the same type.
Furthermore, as reported previously~\cite{Berciaud:04}, we also find
no stringent correlation between the extinction and fluorescence
signals of individual dots.

\begin{figure}[b!]
\centering
\includegraphics[width=8 cm]{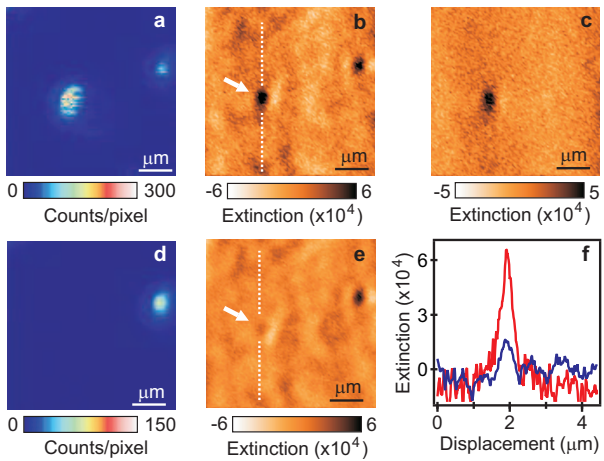}
\caption{\textbf{Photobleaching study} Fluorescence (a) and
extinction (b) images of two individual D2 quantum dots. The marked
dot was the same as the one studied in Fig.~\ref{detection}. (c)
Difference between images (b) and (e). (d,e) Images obtained after
illuminating the marked NQD with 20 kW/$\rm cm^2$ for several
minutes. (f) Cross section of the quantum dot before (red) and after
(blue) photobleaching. }\label{bleaching}
\end{figure}

In addition to pushing the limits of optical imaging,
fluorescence-free detection of emitters provides important insight
into their complex photophysical properties such as photobleaching
and fluorescence blinking. Figures~\ref{bleaching}a and b depict
fluorescence and extinction images obtained from averaging 6
individual scans. Subsequent to the acquisition of these images, the
marked dot was selectively irradiated with an elevated intensity
until it was photobleached and no further fluorescence was observed.
The same area as in (a) and (b) was then scanned to obtain images
(d) and (e), respectively. A comparison of Figs.~\ref{bleaching}b
with e and examination of Fig.~\ref{bleaching}c show that aside from
the marked NQD all image features are reproduced in (e). In fact, it
turns out that the marked dot is also faintly visible in the
extinction image (e) although it is completely absent in the
fluorescence signal (d). Figure~\ref{bleaching}f displays cuts from
Figs.~\ref{bleaching}b and e, revealing a three-fold drop in the
extinction cross section of the NQD. This finding is consistent with
the hypothesis that the NQD polarizability might be decreased as a
consequence of photo-oxidation~\cite{Peng:97}. These results
establish the first report on the interrogation of a single emitter
after its quantum transitions were irreversibly modified.

\begin{figure}[b!]
\centering
\includegraphics[width=8 cm]{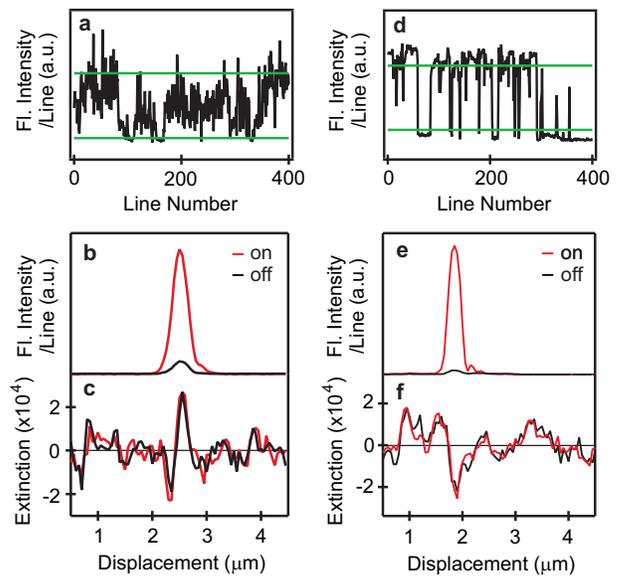}
\caption{\textbf{Blinking study} (a) Sum of fluorescence counts
obtained from one-dimensional line scans across a single D2 NQD for
400 lines. The green lines indicate boundaries where the dot is
determined to be on or off. (b) Resulting fluorescence cross
sections after averaging about 100 lines that lie below and above
the indicated thresholds. (c) Corresponding extinction line scans.
(d-f) Same as (a-c) but for a D1 NQD.}\label{blinking}
\end{figure}

The most fascinating photophysical property of NQDs is their
fluorescence blinking~\cite{nirmal:96}, and many groups have studied
this effect by analyzing the fluorescence
signal~\cite{Kuno:01,Verberk:03,Fisher:04,Pelton:07}. We now show
that our detection mechanism makes it possible to examine the system
even in the absence of fluorescence, i.e. in the blinking off-state.
Following the procedure described earlier, we first identified a
single NQD and then repeatedly scanned it through the laser beam
along one line to record both its fluorescence and extinction data.
Next, we summed the total fluorescence counts for each line and
plotted the outcome for 2500 lines. As shown in
Fig.~\ref{blinking}a, this yields a familiar blinking trace. By
setting lower and upper thresholds shown by the horizontal lines, we
isolated line scans in which the NQD was clearly in the off and on
states. The data selected in this fashion were averaged to arrive at
the fluorescence and extinction line scans shown in
Figs.~\ref{blinking}b and c, respectively. While there is roughly a
10-fold difference in the fluorescence intensity between the on and
off states, no difference in the extinction signal can be observed.
We also repeated the same experiment with D1 NQDs (see
Fig.~\ref{setup}b), which show longer off-times. Consequently, as
plotted in Fig.~\ref{blinking}e, the contrast in fluorescence
between the on and off states is now much stronger. Nevertheless,
Fig.~\ref{blinking}f shows that again the extinction signals of the
on and off states remain identical. We thus conclude that within the
accuracy of our measurement, the room temperature extinction cross
section of a NQD does not change upon blinking.

Although a theoretical treatment of the extinction cross section
away from a sharp resonance is not
straightforward~\cite{Leatherdale:02}, we can gain insight into some
of its fundamental features by considering the cross section of a
two-level quantum system given by
$\sigma=\frac{3\lambda^2}{2\pi}\frac{\gamma_{\rm rad}}{\gamma_{\rm
hom}}$~\cite{Loudon}. Here $\lambda$ is the transition wavelength,
and $\gamma_{\rm rad}$ and $\gamma_{\rm hom}$ denote the radiative
and homogeneous linewidths respectively, whereby $\gamma_{\rm
hom}>\gamma_{\rm rad}$ by 4-5 orders of magnitude at room
temperature. Thus the lack of change in the extinction signal of
NQDs indicates that most likely, neither of $\lambda$,
$\gamma_{rad}$, and $\gamma_{hom}$ has been modified between the on
and off states. In other words, the NQD has not undergone a
transition to a new quantum state. Instead, we believe our
observation is consistent with models where the fluorescence is
quenched in the off state by fast nonradiative relaxation of the
excited state~\cite{Fisher:04,Frantsuzov:05}. To illustrate this, we
point out that $\gamma_{\rm hom}=\sum{\gamma_{\rm i}}$ with
$\gamma_{\rm i}$ signifying various contributions to line
broadening. Hence, if a nonradiative quenching rate
$\gamma_{q}\ll\gamma_{\rm hom}$ is activated, its influence on
$\sigma$ remains negligible while it crucially diminishes the
fluorescence quantum efficiency
$\eta=\frac{\gamma_{rad}}{\gamma_{rad}+\gamma_{q}}$.

The unprecedented detection sensitivity of the extinction
measurements reported in this letter can be improved even further by
several measures. First, the residual optical roughness of the
sample, evident in the comparison of Figs.~\ref{bleaching}b, e, and
c, can be eliminated. In addition, where applicable techniques such
as polarization modulation can help discriminate against such
background fluctuations. Thus, given that the extinction cross
section of a NQD is only a few times larger than that of
conventional dye molecules, extinction detection of these should be
also within reach. In particular, emitters with quenched
fluorescence, for example close to metallic surfaces or in chemical
contact~\cite{Lakowicz-book}, are not disadvantaged in extinction
measurements and become accessible to optical investigations.
Furthermore, by employing white light confocal
microscopy~\cite{Lindfors:04}, it should be possible to record
extinction spectra of individual nanoparticles and molecules in an
efficient manner. These techniques provide exciting opportunities
for extending the applicability of nano-optical studies to a much
wider range of material and environments than has been available to
fluorescence measurements.

The authors thank J. Hwang, M. Agio, N. Mojarad, and G. Zumofen for
discussions and S. Walt for help with the investigation of mica
surfaces. This work was supported by the ETH Zurich, Swiss National
Foundation (SNF) and the Swiss Ministry of Education and Science (EU
Integrated Project Molecular Imaging).

\newpage

\textbf{Methods}

The output of a low noise diode-pumped solid-state laser at 532 nm
is spatially filtered, expanded by a 5:1 telescope, circularly
polarized and split by a 50:50 beam splitter. One half of the light
is sent to a home-built inverted microscope and focused onto the
sample with a high numerical aperture microscope objective (Zeiss
Apochromat, 1.4 NA, 63x). The fluorescence is separated from the
excitation light by a dichroic mirror and a long pass filter before
it is detected by an avalanche photodiode (APD) in the photon
counting mode. The extinction signal is sent through a confocal
pinhole and then a photodiode (PD1) coupled to a low noise current
to voltage amplifier. An iris is used to remove the component of the
excitation light that underwent total internal reflection at the
substrate/air interface. A second amplified photodiode (PD2)
monitors the laser intensity.

Fluorescence and extinction images were acquired by raster scanning
the sample across the focus with a x-y piezo translation stage. By
normalizing and subtracting the signals on both photodiodes, laser
intensity fluctuations are reduced by roughly an order of magnitude
to about $0.01\%$~RMS after normalization. Rapid scanning aided in
reducing noise due to slow drifts in the objective/sample distance
and laser pointing fluctuations.

We used commercially available PEG coated core/shell quantum dots
with emission centered at 565 nm (D1: Invitrogen, Qdot 565 ITK amino
(PEG), CdSe/ZnS), 600 nm (D2: Evident Technologies, EviTags
E2-C11-CB2-0600, CdSe/ZnS) and 680 nm (D3: Evident Technologies,
EviTags E2-C11-NF2-0680, InGaP/ZnS). The stock solutions were
diluted 10000-fold before spin casting 10 $\mu$L of the diluted
solution onto the sample substrate at 3000 rpm for 30 seconds.

We used freshly cleaved mica sheets as substrate to minimize
background fluctuations in the extinction signal~\cite{Jacobsen:06}.
To avoid complications in the imaging caused by the substrate
birefringence, we produced sheets less than $1~\mu$m thick. These
were adhered to standard microscope cover glass with a small drop of
immersion oil for index matching. The sample was immediately placed
on a spin coater and rotated at 3000 rpm for 30 seconds to remove
any excess oil.

The scattering phase $\phi_{\rm sca}$ can be obtained using the
textbook expression for $\alpha$~\cite{Bohren-83book} and the
complex refractive indices of the material of the particle under
study. We used the literature values for the complex refractive
index of the various materials (E. D. Palik, Handbook of Optical
Constants of Solids, (Academic Press, Boston 1985)), as well as an
average refractive index of 1.31 for the mica/air interface to
arrive at scattering phases $\phi_{\rm Au}=0.26\pi$, $\phi_{\rm
CdSe}=0.06\pi$ and $\phi_{\rm InGaP}=0.015\pi$. If we approximate
the focused laser beam as a Gaussian beam, its Gouy phase variation
is given by $\phi_{\rm exc}=tan^{-1}(\lambda z/\pi w_0^2)$, where
$z$ is the displacement from the focus, and $w_0$ is the half beam
waist. The z-coordinate of the focus was determined by fitting the
fluorescence spot to a Gaussian and locating the position at which
the full width at half maximum is minimized.

\end{document}